The Physical Foundation of Human Mind
and a New Theory of Investment

Jing Chen
School of Business
University of Northern British Columbia
Prince George, BC
Canada V2N 4Z9
Phone: 1-250-960-6480
Fax:   1-250-960-5544
Email: chenj@unbc.ca
Web:   http://web.unbc.ca/~chenj/

First draft: February, 2003
This version: December, 2005

We thank David Faulkner, Andreas Herzog, Frank Schmid, Dietrich Stauffer and participants of MFA and FMA conferences for helpful comments. The paper was previously titled The Physical Foundation of Human Psychology and Behavioral Finance.

Abstract

This paper consists of two parts. In the first part, we develop a new information theory, in which it is not a coincidence that information and physical entropy share the same mathematical formula. It is an adaptation of mind to help search for resources. We then show that psychological patterns either reflect the constraints of physical laws or are evolutionary adaptations to efficiently process information and to increase the chance of survival in the environment of our evolutionary past. In the second part, we demonstrate that the new information theory provides the foundation to understand market behavior. One fundamental result from the information theory is that information is costly. In general, information with higher value is more costly. Another fundamental result from the information theory is that the amount of information one can receive is the amount of information generated minus equivocation. The level of equivocation, which is the measure of information asymmetry, is determined by the correlation between the source of information and the receiver of information. In general, how much information one can receive depends on the background knowledge of the receiver. The difference in cost different investors are willing to pay for information and the difference in background knowledge about a particular information causes the heterogeneity in information processing by the investment public, which is the main reason of the price and volume patterns observed in the market. Many assumptions in some of the recent models on behavioral finance can be derived naturally from this theory.

The proliferation of new investment theories suggests that a more fundamental approach is necessary. In this paper, we investigate the physical foundation of human mind and develop a new information theory that provides a unified understanding of information and physical entropy. This new information theory provides the foundation to understand major patterns in asset markets.

There is a growing consensus that security markets are not as efficient as we thought before. The inefficiency is generally attributed to behavioral biases of investors. Since many behavioral biases have been documented in the cognitive psychology literature, almost any patterns in the financial markets can be linked to one or several of these biases. However, "the potentially boundless set of psychological biases that theorists can use to build behavioral models and explain observed phenomena creates the potential for 'theory dredging.'" (Chan, Frankel and Kothari, 2002) Furthermore, many theories, while consistent with empirical patterns that they are set out to explain, are not consistent with other empirical results. For example, while Bloomfield and Hales (2002) find evidence supportive of behavioral model of Barberis, Shleifer, and Vishny (1998) in a laboratory setting, Durham, Hertzel and Martin (2005) find scant evidence that investors behave in accordance with the model using market data. The link between behavioral theory and investment behavior are often vague. For example, empirical works reveal that small investors' trading activities often hurt their investment return (Hvidkjaer, 2001). This is usually thought that small investors are more prone to behavioral biases than professionals, who are better trained (Shanthikumar, 2004). Yet some empirical work suggests that professionals exhibit some behavioral biases to a greater extent than non-professionals (Haigh and List, 2005).

Behavioral models often assume some market sectors or activities are irrational while other parts are rational. However, they often differ on which part should be considered irrational. For example, Shleifer and Vishny (2003) develop a new model on corporate mergers. "Mergers in this model are a form of arbitrage by rational managers operating in inefficient markets. This theory is in a way the opposite of Roll's (1986) hubris hypothesis of corporate takeovers, in which financial markets are rational, but corporate mangers are not. In our theory, managers rationally respond to less-than rational markets" (Shleifer and Vishny, 2003, p. 297).

In a recent work, Ivkovic and Weisbenner (2005) questioned whether investor performance should be interpreted primarily as a behavioral phenomenon or information-driven. They conclude, at least in their sample, that information is the primary reason of investment performance. Empirical results also suggest that superior performance of investors trading local securities and geographically proximate analysts is due to an information advantage over others (Coval and Moskowitz, 2001; Malloy, 2005). Kacpercyzk, Sialm and Zheng (2005) documented that, mutual funds whose portfolios are highly concentrated in industries where the fund managers have informational advantages perform better than mutual funds with well-diversified portfolios. The empirical evidences suggest we should explore an information based theory of market behavior. However, "theory offers little guidance in identifying informed investors and in distinguishing between securities with scarce information and those with widely available information" (Coval and Moskowitz, 2001).

The standard economic theory of information was developed by Grossman and Stiglitz (1980). This theory is based on rational expectation. It assumes that investors can accurately assess the value of some information and pay some fixed amount accordingly to obtain the information. Recently, various models relax the rational expectation assumption to explain major market patterns. Most of these models rely on some kind of human psychological biases. The flourishing of behavioral models and interpretations suggests that a more fundamental theory is needed.

The basic problem of Grossman and Stiglitz theory is that it does not actually model how human beings process information. More than fifty years ago, Shannon (1948) developed the entropy theory of information. The major advantage of Shannon's theory is that it can determine how much information a communication system can receive from the source of information. Recently, a new theory of information was developed to expand Shannon's entropy theory of information into an economic theory (Chen, 2003, 2005). The basic idea of this theory is to assert that information is the reduction of entropy, not only in a mathematical sense, as in Shannon's theory, but also in a physical sense. The physical cost of information is highly correlated with economic cost of information. The rules of information transmission developed in Shannon's theory, as mathematical rules, apply not only to communication systems, but also to all living organisms, including human beings. In this work, we will show how this new information theory can offer "guidance in identifying informed investors and in distinguishing between securities with scarce information and those with widely available information" and provides a unified understanding of observed market behaviors documented in increasingly abundant empirical works.

Several important properties can be derived from this new information theory. First, information with higher value is in general more costly. This is a direct extension from Maxwell's (1871) thought experiment on an intelligent demon. Second, the amount of information one can receive is the amount of information generated minus equivocation. The level of equivocation, which is the measure of information asymmetry, is determined by the correlation between the source of information and the receiver of information. In general, how much information one can receive depends on the background knowledge of a person. Therefore the process of understanding information is a process of learning, which often takes long time. This result is a direct extension from Shannon's theory from technical communication systems to human cognitive systems. Third, the value of information is inversely related to the number of people who understand it. For example, an investor who buys the shares of a company before it becomes popular often earns higher rate of return than those who buy the shares of the same company when it becomes hot. Buffet (2001) once commented, "What the few bought for the right reason in 1925, the many bought for the wrong reason in 1929." At the same time, the value of a company's investment is also affected by how much its competitors understand the technology and market potential of a product.

After the new information theory is developed, the new theory of investment can be simplied stated as the following:

> Major patterns in asset markets are the result of information processing by the heterogeneous investment public.

Later, we will discuss detailed theoretical predictions and empirical evidences of this theory.

As the efficient market theory retreats, many new theories have emerged and more new theories will emerge to fill the void left behind. With so many new theories around, it is time to consider criteria for a good theory.

First, a good theory should be consistent with empirical evidences. Not so long ago, empirical works on market behavior were so few that it was very difficult to distinguish the validity of different theories. In the past several years, however, we witness the emergence of a growing number of empirical works, which often call into the question of the ability of existing behavioral theories in explaining broader sets of empirical patterns. In this work, we will list these empirical results and show how the new information theory can provide a simple and unified understanding of the empirical patterns. This theory will resolve some puzzles about the market patterns raised in the recent literature that could not be answered by existing theoretical frameworks.

Second, a good theory should provide more precise predictive power than the existing theories. Kahneman and Tversky's prospect theory is an improvement over Simon's bounded rationality theory because it offered some concrete patterns of irrationality beyond the general statement of bounded rationality. Empirical evidences indicate that the patterns of trading by small, individual investors differ systematically from that by large, institutional investors. Yet the existing behavioral theories do not offer any particular behavioral explanations to this systematic difference (Hvidkjaer, 2001; Shanthikumar, 2004). The new information theory, however, provide very precise understanding of the systematic differences between trading patterns of large and small investors. This information theory states that more valuable information is more costly to obtain in general. For large investors, it pays to spend a lot of effort and money to research the fundamentals. For small investors, it doesn't pay to dig into the fundamentals. They depend on processed and easy to understand information that is readily available at low cost, such as news from popular media and price movement of the shares. In this work, we will show that patterns of returns by small and large investors and patterns of return and trading volumes of stocks can be easily understood from this information processing perspective.

Third, a good theory should help provide deeper understanding to existing models. For example, Hong and Stein's (1999) results are built on three key assumptions. The first two assumptions are that traders are classified as "newswatchers" and "momentum traders" according to their information processing abilities. They commented that, "the constraints that we put on traders' information-processing abilities are arguably not as well-motivated by the experimental psychology literature as the biases in Barberis et al. (1998) or Daniel et al. (1998), and so may appear to be more ad hoc" (Hong and Stein, 1999, p. 2145). These assumptions can actually be derived naturally from this new information theory. Depending on the value of assets under management, different investors will choose different methods of information gathering with different costs. "Newswatchers" are large investors who are willing to pay a high cost to collect private information and to make a deep understanding of public information. "Momentum traders" are investors who spend less cost or effort on information gathering and rely mainly on easy to understand low cost information such as coverage from popular media and price momentum signals. Cohen, Gompers and Vuolteenaho (2002) empirically confirm that institutional investors buy on fundamental news while individual investors buy on price trends. The third assumption of Hong and Stein (1999) is that private information diffuses gradually across the newswatcher population. The gradual diffusion of private information means that people gradually learn about the background knowledge of information and understand information better over time. This new information theory provides clear understanding to all three assumptions in Hong and Stein's model.

Empirical evidences show that patterns in the security markets are mainly driven by patterns of information processing by the investment public. However, this does not mean that human emotions do not influence market behaviors. It has been shown that most psychological patterns either reflect the constraints of physical laws or evolutionary adaptation to process information more efficiently in our evolutionary past (Chen, 2003, 2005). In a word, human emotions are really low cost, but not necessarily unbiased ways of information processing. By showing that most patterns in asset markets documented in the literature can be explained by this information theory, the focus of attention can be directed to search the links between human biases and the phenomena or magnitude of phenomena in asset markets that could not be explained by the information theory. Furthermore, human activities, including mental activities, are constrained by physical laws. These constraints offer initial tests to the plausibility of assumptions in behavioral theories. For example, from the information theory, new information can only be understood gradually by human beings. If a behavioral theory suggests investors will generally overact to new information, we need to examine the empirical evidence with great caution.

This paper is an update from earlier works Chen (2003, 2004, 2005, 2006). The remainder of the paper is structured as follows. Section I presents the generalized entropy theory of information. This information theory provides natural measures of the cost of obtaining information and of information asymmetry. It offers a deep connection among information, entropy and human mind. Section II shows that entropy theory offers a unified understanding of the patterns of human psychology and explains why successful investors in one era often perform worse than novice when new industry emerges. Section III explains how patterns of investor returns and patterns of security returns and trading volumes are natural results of information processing by the heterogeneous investment public. It also answers many questions on the gaps between existing behavioral theories and empirical evidences. Section IV discusses how this information theory based model of investor behavior is related to other models of behavioral finance. Many assumptions in some of the recent theoretical models can be derived naturally from the generalized entropy theory of information. Section V concludes.

**I. Entropy, resource, information, and the evolution of mind**

The value of information is a function of probability and must satisfy the following properties:

(a) The information value of two events is higher than the value of each of them.
(b) If two events are independent, the information value of the two events will be the sum of the two.
(c) The information value of any event is non-negative.

The only mathematical functions that satisfy all the above properties are of the form

$$H(P) = -\log_b P \tag{1}$$

where $H$ is the value of information, $P$ is the probability associated with a given event and $b$ is a positive constant (Applebaum, 1996). Formula (1) represents the level of uncertainty. When a signal is received, there is a reduction of uncertainty, which is information.

Suppose a random event, $X$, has $n$ discrete states, $x_1, x_2, \ldots, x_n$, each with probability $p_1, p_2, \ldots, p_n$. The information value of $X$ is the average of information value of each state, that is

$$H(X) = -\sum_{j=1}^{n} p_j \log(p_j) \tag{2}$$

The right hand side of (2), which is the entropy function first introduced by Boltzmann in the 1870s, is also the general formula for information (Shannon, 1948).

After the entropy theory of information was developed in 1948, its technique has been applied to many different problems in economic and finance. (Theil, 1967; Maasoumi and Racine, 2002 and many others) However, the standard economic theory of information, represented by Grossman and Stiglitz (1980) was not built on the foundation of entropy theory. Economists often feel that "the well-known Shannon measure which has been so useful in communications engineering is not in general appropriate for economic analysis because it gives no weight to the value of the information. If beforehand a large manufacturer regards it as equally likely whether the price of his product will go up or down, then learning which is true conveys no more information, in the Shannon sense, than observing the toss of a fair coin." (Arrow, 1973 (1983), p. 138)

The Shannon measure actually carries weight of information. For example, $N$ symbols with identical Shannon measure carry $N$ times more information than a single symbol (Shannon, 1948). Similarly, the value of the information about the future price is higher to a large manufacturer than to a small manufacturer, other things being equal.

Authorities in information theory also discouraged the application of entropy theory to broader areas.

> Workers in other fields should realize that that the basic results of the subject are aimed at a very specific direction, a direction that is not necessarily relevant to such fields as psychology, economics, and other social sciences. Indeed, the hard core of information theory is essentially, a branch of mathematics, a strictly deductive system. (Shannon, 1956)

This orthodox view was reaffirmed recently:

> The efforts of physicists to link information theory more closely to statistical physics were less successful. It is true that there are mathematical similarities, and it is true that cross pollination has occurred over the years. However, the problem areas being modeled by these theories are very different, so it is likely that the coupling remains limited.
>
> In the early years after 1948, many people, particularly those in the softer sciences, were entranced by the hope of using information theory to bring some mathematical structure into their own fields. In many cases, these people did not realize the extent to which the definition of information was designed to help the communication engineer send messages rather than to help people understand the meaning of messages. In some cases, extreme claims were made about the applicability of information theory, thus embarrassing serious workers in the field. (Gallager, 2001, p. 2694)

However, the dissonance between entropy function as a mathematical representation of information and the practical value of information has long puzzled many people and recent works have shown that our intuitive concept of information coincides with the mathematical definition of information as entropy (Bergstrom and Lachmann, 2004; Adami, 2004). In the following we will provide a more formal argument. If some decision making process is truly important and is needed again and again in life, it is highly economical that quantitative modules to be evolved in the mind to expedite the process. For example, predators need routinely to assess their distance from the prey, the geometry of the terrain, the speed differential between itself and the prey, the energy cost of chasing down its prey, the probability of success of each chase and the amount of energy it can obtain from prey to determine whether, when and where to initiate a chase. There are many other sophisticated functions, such as navigation by migrating birds over long distances, that need sophisticated mathematical capabilities. Many animals need to make precise calculations of many of these quantitative problems many times in life. To reduce the cost of estimation, mathematical models must have evolved in their mind so that many decision making processes are simplified into parameter estimation and numerical computation. It is highly likely that, if some function is very important for the survival of the animal, in the process of evolution, this function will be genetically assimilated. The nature of the entropy function may be gleaned from the following quote:

> The German physicist Rudolf Clausius coined the word "entropy" in 1865. Looking for a word similar to the word "energy", Clausius chose the Greek word "entropy", which means (in Greek) "the turning" or "the transformation." The coinage is indeed apt: if we ask, "which transformation of a system will occur spontaneously?", then the multiplicity of each of the various possible macrostates is crucial, for the system will evolve to the macrostate of largest multiplicity. (Baierlein, 1999, p. 34)

More generally, a system has a tendency to move from a state of smaller multiplicity (or probability) to a state of larger multiplicity (or probability). This tendency of directional movement is the source of useful energy that drives, among other things, the living organisms. Since all living organisms need to extract low entropy from the environment to compensate continuous dissipation and entropy is the only mathematical function to measure scarcity of resources (Chen, 2005), it is inevitable that information, which we collect for our survival, is largely about entropy. It is not a mere coincidence that our intuitive concept of information and the mathematical definition of information as entropy largely overlap. The intuitive concept is really a simplified evaluation of a mathematical computation.

If Shannon's entropy theory of information is purely a mathematical theory with little connection with the physical laws, it would be a miracle that information defined as entropy turns out to have the magic properties that handle technical problems in communication so well. However, once mathematical theories are thought to be a natural part of our evolutionary legacy, it would be natural for entropy theory of information to possess this property.

From the above discussion, an entropy theory based economic theory of information can be simply stated as:

> Information is the reduction of entropy, not only in a mathematical sense, as in Shannon's theory, but also in a physical sense. The rules of information transmission developed in Shannon's theory, as mathematical rules, apply not only to communication systems, but also to all living organisms.

In the following, we will discuss some distinct properties of this new information theory. First, information that is more valuable is in general more expensive to obtain. From the second law of thermodynamics, Maxwell (1871) concluded that information of higher value is of higher physical cost. Since economic cost is highly correlated to physical cost (Georgescu-Roegen, 1971; Chen, 2005), more valuable information is in general more expensive to obtain.

Second, the amount of information one can receive depends on the person's background knowledge about that particular information. The most important result from Shannon's entropy theory of information is the following formula

$$R = H(x) - H_y(x) \qquad (3)$$

where $R$ is the amount of information one can receive, $H$ is the amount of information a source sent and $H_y(x)$, the conditional entropy, is called equivocation. Formula (3) shows that the amount of information one can receive would be equal to the amount of information sent minus the average rate of conditional entropy. Before Shannon's theory, it was impossible to accurately assess how much information one can receive from an information source. In communication theory, this formula is used to discuss how noises affect the efficiency of information transmission. But it can be understood from more general perspective. The level of conditional entropy $H_y(x)$ is determined by the correlation between senders and receivers. When $x$ and $y$ are independent, $H_y(x) = H(x)$ and $R = 0$. No information can be transmitted between two objects that are independent of each other. When the correlation of $x$ and $y$ is equal to one, $H_y(x) = 0$. No information loss occurs in transmission. In general, the amount of information one can receive from the source depends on the correlation between the two. The higher the correlation between the source and receiver, the more information can be transmitted.

The above discussion does not depend on the specific characteristics of senders and receivers of information. So it applies to human beings as well as technical communication equipments, which are the original focus in information theory in science and engineering. However, the laws that govern human activities, including mental activities, are the same physical laws that govern non-living systems.

$H_y(x)$ in Formula (3) offers the quantitative measure of information asymmetry (Akerlof, 1970). Since different people have different background knowledge about the same information, heterogeneity of opinion occurs naturally. To understand the value of a new product or new production system may take the investment public several years. To fully appreciate the scope of some technology change may take several decades. For example, the economic and social impacts of cars as personal transportation instruments and computers as personal communication instruments were only gradually realized over the path of several decades. This is why individual stocks and whole stock markets often exhibit cycles of return of different lengths. This property is very different from Grossman-Stiglitz information theory, where economic agents can recognize the value of information instantly and pay according to its value.

Third, the same information, when known to more people, becomes less valuable. Figure 1 is a graph of (1), where $H$ is a function of $P$, the probability of any given event. From Figure 1, value is a decreasing function of probability. In the standard information theory, $P$ represents the probability that some event will occur. In this theory, $P$ is generalized to represent the percentage of people or money that is controlled by informed investors. When $P = 1$, $-\log P = 0$. Thus the value of information that is already known to everyone is zero. When $P$ approaches zero, $-\log P$ approaches infinity. Therefore, the value of information that is known to few is very high. The following example will illustrate this point. Figure 2 shows overnight rate of return and trading volume of shares of WestJet, a Canadian airline, surrounding the announcement of the bankruptcy of Jetsgo, the main competitor of WestJet. Jetsgo announced bankruptcy at the evening of March 10, 2005. If one bought stock at March 10, he would have made a return of 40% overnight. Judging from the trading volume of March 9, some people did buy WestJet stock before information was released to the public. After the announcement made the information public, trading volume was very high and the rate of return is near zero. Figure 2 neatly illustrates the relation between value of some information and the number of people who know the information.

It is often said that the cost of information has dropped sharply over the years. But at the same time, the value of the same type of information has dropped sharply as well. Information of high value is usually carefully guarded and difficult to detect. For example, Warren Buffett, who has a very successful record for gaining and using insightful market information, would not announce to the public which stock(s) he is going to buy or sell. Animals have discovered this long ago. "In those cases where animal signals really are of mutual benefit, they will tend to sink to the level of a conspiratorial whisper: indeed this may often have happened, the resulting signals being to inconspicuous for us to have noticed them. If signal strength increases over the generations this suggests, on the other hand, that there has been increasing resistance on the side of the receiver." (Dawkins, 1999, p. 59)

Unlike Grossman-Stiglitz information theory, this information theory is a non-equilibrium theory. It does not assume a company possesses some intrinsic value waiting to be discovered by the investment public. Instead, the process of understanding the value of a company by the investment public is accompanied by the process of understanding the technology and market potential by its competitors, which generally reduce the value of that particular company. Empirical evidences that we will present later support this statement.

Since this new information theory can be applied to much broader fields than Shannon's theory, it may be called the Generalized Entropy Theory of Information. Empirical evidences show that Grossman-Stiglitz information theory cannot explain observed investor behaviors (Barber and Odean, 2000). Other models extended from Grossman-Stiglitz theory often have little explanatory power outside a limited scope. We will show later that the generalized entropy theory of information will provide a unified understanding of the empirical market behaviors documented in the literature.

## II. The entropy theory of human psychology

Human activities, as well as the activities of all organisms, are essentially thermodynamic processes, or entropic processes. (Schrodinger, 1944; Wiener, 1948; Prigorgine, 1980)

The entropy law, which states that closed systems tend towards states of higher entropy, is the most universal law of the nature. From the entropy law, we know that it is far easier for a system to disintegrate than to maintain its structure. (Morowitz, 1992; Margulis, 1998) So there is a strong selective pressure for important knowledge to become genetically coded into heuristic principles to reduce the cost of learning. (Tversky and Kahneman, 1974) Natural selection determines that human minds are born with many data and preferences stored. (Pinker, 1997) In this section we will show that entropy theory offers a unified understanding of some frequently cited patterns of human psychology in behavioral finance literature.

1. Conservatism

Conservatism in human beings may be characterized as behavior by individuals who possess a reluctance to update their beliefs in the face of new information. This property is a natural result from information theory. From (3), the information one can receive is information sent minus equivocation, which is reduced gradually as the receiver's background knowledge about the source increases. Hence conservatism reflects the gradual reduction of equivocation by the receiver of any given information.

This theory provides a clear link between conservatism and Bayesian updating, which we consider rational. From (3), as equivocation approaches to zero, the asymptotic state becomes Bayesian estimate.

2. Framing or representativeness

We often frame, or sort problems into categories and assign different associated values based on the perceived relative levels of importance of each problem. Why do we do this? Apart from the self-evident common sense answer, the following result from statistical physics helps answer this question.

If $\{p_1, \ldots p_n\}$ and $\{q_1, \ldots q_n\}$ are two sets of probabilities, then

$$-\sum_{j=1}^{n} p_j \log(p_j) \leq -\sum_{j=1}^{n} p_j \log(q_j) \qquad (4)$$

with equality achieved if and only if each

$$q_j = p_j, \qquad 1 \leq j \leq n$$

This result is called Gibbs inequality. (Isihara, 1971) In Gibbs inequality, $p_j$ can be understood as the probability of event $j$ in nature and $q_j$ is the subjective probability of our assessment of that event. The left hand side of formula (4) is the average uncertainty of events and the right hand side is the uncertainty of our subjective assessment of those events. In general, the difference between the left hand side and right hand side of (4) is smaller when $q_j$ is closer to $p_j$. This means that information processing is more efficient when the subjective probabilities are closer to the objective probabilities. Such a conclusion makes both common sense and matches our intuition. In particular, a mind with stored data about the natural environment is in general more efficient than a completely unbiased mind, where all subjective probabilities are to be learned from scratch. Natural selection determines that the human mind will evolve so that, "in general, instances of large classes are recalled better and faster than instances of less frequent classes; that likely occurrences are easier to imagine than unlikely ones; and that the associative connections between events are strengthened when the events frequently co-occur." (Tversky and Kahneman, 1974, p.1128)

Thus, "people rely on a limited number of heuristic principles which reduce the complex tasks of assessing probabilities and predicting values to simpler judgmental operations. In general, these heuristics are quite useful, but sometimes they lead to severe and systematic errors." (Tversky and Kahneman, 1974, p.1124) What causes these severe and systematic errors? Human minds are the result of natural selection, which "operates over thousands of generations. For ninety-nine percent of human existence, people lived as foragers in small nomadic bands. Our brains are adapted to that long-vanished way of life, not to brand-new agriculture and industrial civilizations." (Pinker, 1997, p. 42) This is why we observe systematic errors in judgment by human beings, i.e., the typical framework for processing information today was developed over millennia when environmental conditions were very different. For example, most of us still have a great fear of snakes, although they rarely pose a threat to urban dwellers today. On the other hand, fear of electricity has to be instilled into children's minds with great difficulty. (Pinker, 1997)

From Gibbs inequality, the level of uncertainty in understanding a type of events is

$$-\sum_{j=1}^{n} p_j \log(q_j)$$

where $p_i$ and $q_i$ are objective and subjective probabilities respectively. Suppose this type of events has two possible outcomes, state 1 and state 2. The probability of state 1 is 90% and the probability of state 2 is 10%. An expert on this type of events may correctly estimate these probabilities and for her the uncertainty in prediction is

$$-0.9\ln 0.9 - 0.1\ln 0.1 = 0.33$$

A novice, who has no priori knowledge on these events, may assign 50% probability to each outcome. For her the uncertainty in prediction is

$$-0.9\ln 0.5 - 0.1\ln 0.5 = 0.69$$

It is clearly that the expert, who has accumulated knowledge through long time experience, has better estimation than novice in a stable environment.

Now assume the environment experiences some fundamental change and the new probabilities of state 1 and state 2 become 10% and 90% respectively. This time, the uncertainty of the prediction by the expert, who still uses the same probability, is

$$-0.1\ln 0.9 - 0.9\ln 0.1 = 2.08$$

while the uncertainty of prediction by a novice is

$$-0.1\ln 0.5 - 0.9\ln 0.5 = 0.69$$

This shows that when environment changes suddenly, novice actually perform better than experts, whose priori knowledge often cause severe biases in prediction. This is why successful investors cannot eliminate other investors continuously and why financial markets will not become progressively more efficient over time. This also explains why new industries are often pioneered by newcomers or outsiders.

3. Herd behavior

From the second law of thermodynamics, a random action generally costs more than it gains. To concentrate actions into profitable ones, we, like wild animals, often learn from the experience of successful individuals and copy their behavior. It is generally very costly and impossible to repeat all of the experiences and mistakes that are possible. Therefore, we accept certain modes of behavior demonstrated by others without completely investigating the reasons behind them. Copying the actions of others directly is much easier, i.e., more efficient. Herding mentality developed because it is a cost-effective way of learning most of the time. That said, herding can be another reason that systematic errors in judgment are observable by individuals and societies.

4. Overconfidence and irrationality

"Extensive evidence shows that people are overconfident in their judgments." (Barberis and Thaler, 2003) From entropy law, any biological system, as a non-equilibrium system, faces continuous dissipation of energy. Endless efforts are required to maintain a non-equilibrium system. Entropy law has been intuitively understood since ancient times. "The gods had condemned Sisyphus to ceaselessly rolling a rock to the top of a mountain, whence the stone would fall back of its own weight. They had thought with some reason that there is no more dreadful punishment than futile and hopeless labour. … If this myth is tragic, that is because its hero is conscious. … The workman of today works every day in his life at the same tasks and this fate is no less absurd. But it is tragic only at the rare moments when it becomes conscious." (Camus, 1955, p. 109) In the long course of evolution of our solar system, all life on earth will eventually go extinct in the far distant future (Lovelock, 1988). From a purely rational perspective, life is meaningless. Since human beings are self-conscious, the very question of why life is worth living lingers in many people's minds. "There is but one truly serious philosophical problem and that is suicide. Judging whether life is or is not worth living amounts to answering the fundamental question of philosophy." (Camus, 1955, p. 11) Overconfidence and irrationality are adaptive psychological traits that help us survive in this world.

5. Loss aversion in winning and risk seeking in losses

Human beings often exhibit loss aversion in winning and risk seeking in losses. Kahneman and Tversky (1979) collected some responses to hypothetical choice problems. In one problem, the subjects were presented with two choices.

Choice A: There is an 80% probability of winning 4000 pounds and a 20% probability of winning nothing.

Choice B: There is a certainty of winning 3000 pounds.

The expected end wealth of choice A is 3200 and of choice B is 3000. Most respondents chose B, exhibiting loss aversion in winning. When the signs of the outcomes are reversed, the problems become the following:

Choice C: There is an 80% probability of losing 4000 pounds and 20% probability of losing nothing.

Choice D: There is a certainty of losing 3000 pounds.

The expected end wealth of choice C is -3200 and of choice D is -3000. Most respondents chose C, exhibiting risk seeking in losses. As money is a new invention in human evolutionary history, the preference for money must be derived from something else. Since food is the most important resource of our evolutionary past, our preference for wealth is probably derived from our preferences for food.

In the history of human evolution, we have not been able to store large amounts of extra food. If one goes without food for several days, he will starve. We translate the monetary numbers from the above four questions into days of food to obtain the following. In the case of gain, we can think of the choices of two possible strategies. In the first strategy, there is an 80% probability of getting food for 40 days and 20% chance of getting nothing. In the second strategy, there is a certainty of getting food for 30 days. It is easy to see why most people will prefer 30 days of food in certainty over a strategy that contains a 20% risk of getting nothing. In the case of loss, we can think of the choices of two possible strategies. In the first strategy, there is an 80% probability losing food for 40 days and a 20% chance of getting nothing for zero day. In the second strategy, there is a certainty of getting no food for 30 days. Since 30 days' without food will represent sure death, people will naturally choose a 20% chance of survival. So people consistently avoid risk in both positive gain and negative loss. "Risk seeking" in loss is an unfortunate terminology borrowed from utility theory.

From the above discussion, we find that some psychological patterns, such as conservatism, reflect the constraints of thermodynamic laws. Others, such as framing and herding, are evolutionary adaptations to enable efficient processing of information, which is the reduction of entropy. Still others, such as overconfidence, irrationality and loss aversion, are mental attitudes that help us survive the continuous dissipation of energy endured by all non-equilibrium systems. Therefore, entropy theory offers a unified understanding of human psychology.

In the next section, we will apply the generalized entropy theory of information, a theory based on most fundamental physical laws, to understand the patterns in the stock market without directly invoking assumptions on human psychology. This will avoid the problem of overfitting theories to empirical observation.

**III. Theoretical predictions and empirical evidences**

*A. Differences in informational advantages of investors of different sizes*

From this information theory, information with higher value is in general more costly. The amount of time and resource an investor will spend depends on the value of her portfolio. It can be predicted that the rate of return for large, institutional investors will be higher than small, individual investors because large investors will spend more time and resources in research. As the market rate of return is the average rate of return of all investors, we expect large investors outperform market average and small investors underperform market average. This is supported by empirical evidences. (Wermers, 2000; Barber and Odean, 2000; Cronqvist and Thaler, 2004)

If it is not economical for small investors to spend a lot of effort to study individual stocks, can they expect superior returns by investing in mutual funds with talented managers? The information theory predicts that the value generated by the talented managers will be mainly retained by the managers, who spend time to collect and analyze information, and not be distributed to mutual fund investors. Wermers (2000) documented that during the 1977 to 1994 period, mutual funds did generate higher raw rate of return than the market indices, which indicates researches by mutual fund managers do uncover valuable information. However, the net rate of return to investors is an average of 13.3 percent per year, which is the same as the Vanguard fund, the largest index fund. As the average transaction costs decrease over the years, average expense ratios increase, making the sum of these two costs remaining relatively constant over the years. (Wermers, 2000) This shows that it is the mutual fund managers, not the average investors, who capture the benefit of declining transaction costs. This is consistent with the model of Berk and Green (2004), where mutual fund managers capture the entire rents from their performance. In essence, for small investors without special information advantage or people whose cost of collecting and analyzing information outweighs its benefit, which are the majority of the population, they cannot expect a higher rate of return on stocks than the general market. The above result may be called the generalized efficient market hypothesis: investors without informational advantage can not outperform the general market.

Odean (1999) documented that the shares individual investors sold outperform the shares they bought. He attributed this to some unknown biases of investors' judgment and stated, "What is more certain is that these investors do have useful information which they are somehow misinterpreting." (Odean, 1999, p. 1296) The real reason may be a lot simpler. Chen, Jegadeesh and Wermers (2000) documented that shares bought by mutual fund managers outperform shares they sold. The differential performance of shares bought and sold by individual investors is partly due to informational advantages of some of their trading counterparties. Even if individual investors randomly select buy and sell orders, most trades that get executed are not in their favor, because some of their counterparties, such as mutual fund managers, possess valuable information and select to fill the orders that are in their favor. For example, if an individual investor randomly select two stocks to sell with limit orders, it is more likely that the stock which will have a higher rate of return in the future will be bought by more informed investors. Later, we will demonstrate that small investors' dependence on low cost, low value information does make them prone to bad trading decisions.

Barber and Odean (2000) documented that there is very little difference in the gross performance of households that trade frequently and those that trade infrequently. On average, shares individual investors sold outperform the shares they bought. (Odean, 1999) If all else are same, more frequently traded investors should earn lower rate of gross return. So this indicates that more frequently traded investors have better investment skills than less frequently traded investors. This prediction can be directly verified from trading records. Barber and Odean cited Carhart (1997) to show that frequent trading also hurts the performance of mutual funds. However, a more recent and more detailed study by Wermers (2000) shows that the most traded mutual funds outperform least traded mutual funds by a wide margin. Barber and Odean attribute trading activities to overconfidence, a behavioral explanation. It may be simpler and more consistent with empirical evidences to explain trading as a type of learning activity. Like all other kinds of learning, trading is costly. The cost of learning is compensated, at least partially, by knowledge gained from the experience.

*B. Differences in informational advantages of investors of different backgrounds*

From (3), the ability to understand information depends on the background knowledge of investors. This indicates that investors will earn higher rate of return if they choose to invest in securities that they are familiar with. This is supported by some recent empirical investigations. Professional managers' and individual investors' local investments outperform their remote investments. (Coval and Moskwitz, 2001; Ivkovic and Weisbenner, 2005) Malloy (2005) provides evidence that local analysts are significantly more accurate than other analysts in forecasting and recommendations. Mutual funds with high industry concentration, where fund managers can focus on particular industries they are familiar with, are more successful in selecting securities than diversified funds. (Kacpercyzk, Sialm and Zheng, 2005) Investors take longer time to understand information from sources they are less familiar with. Hong, Lim and Stein (2001) empirically confirm that information from small firms, from firms with low analyst coverage and from firms with bad news, which managers are reluctant to release, generally diffuse slower. From Hvidkjaer (2001), the selling pressures on losers generally are stronger and last much longer than buying pressures on winners, suggesting information processing is less efficient on bad news.

Since insiders understand information much better than others, they can and will take advantage of this information asymmetry to pursue certain corporate activities. Shleifer and Vishny (2003) develop a new model on corporate mergers, in which mergers are a form of arbitrage by rational managers operating in inefficient markets. They show that their model has better explanatory power than the behavioral based models and is supported by new empirical evidences. (Ang and Zheng, 2002) While assuming managers are rational, they continue to assume financial markets are less-than-rational. However, form the information theory, there is no need to assume financial markets are less-than-rational. In the next subsection, we will show how information processing by investors of different sizes with different background knowledge will generate cycles of undervaluation and overvaluation.

*C. Information processing by heterogeneous investors and market patterns*

A persistent pattern in the security market is the price continuation in short to medium run and the reversal of return in the long run. (DeBondt and Thaler, 1985; Jegadeesh and Titman, 1993) Several models have been developed to explain this pattern. (Barberis, Shleifer and Vishny,1998; Daniel, Hirshleifer and Subrahmanyam,1998; Hong and stein,1999) However, these models are based on some ad hoc assumptions and could not explain other patterns. (Lee and Swaminathan, 2000; Hvidkjaer, 2001) For example, the return patterns are often accompanied by distinct patterns of trading volume. However, "existing theories of investor behavior do not fully account for all of the evidence. … none of these models incorporate trading volume explicitly and, therefore, they cannot fully explain why trading volume is able to predict the magnitude and persistence of future price momentum." (Lee and Swaminathan, 2000, p. 2066)

Lee and Swaminathan (2000) and Hvidkjaer (2001) presented many empirical results regarding return, trading volume and patterns of investment decisions by traders of different sizes. Many of these results, especially those in Hvidkjaer (2001), are not presented in a systematic way because of the lack of a proper theoretical framework. In the following, we will show how the generalized entropy theory of information offers a coherent and comprehensive understanding of observed market patterns and the underlying dynamics. Hvidkjaer's work has gone through several significant revisions.

Our discussion will be based on his new version from July, 2004. This part is mainly adapted from Chen (2004).

To an investor, the choice of information gathering is a matter of cost and the ability to understand some information depends on her background. More valuable information is more costly to obtain in general. For large investors, it pays to spend a lot of effort and money to research the fundamentals. For small investors, it doesn't pay to dig into the fundamentals. They depend on processed and easy to understand information that is readily available at low cost, such as news from popular media and price movement of the shares. Cohen, Gompers and Vuolteenaho (2002) confirmed that institutional investors buy on fundamental news while individual investors buy on price trends. Whether an investor will be able to understand certain information also depends on her particular background, which determines her level of equivocation in receiving that information. Information processing by investors of different sizes with different backgrounds will generate the return and trading volume cycle that is similar to Lee and Swaminathan's (2000) momentum life cycle. In the following, we will illustrate the patterns of return and trading volumes of a stock of a typical company from the information processing perspective.

Suppose a company develop a new technology, which is expected to bring the company high profit in the future. From (3), those who are familiar with the technology and company will have low equivocation in receiving the information. They understand the significance of the information first and buy the company shares. Since they are a small number of people, the buying is of low volume. This corresponds to the beginning of the low volume winner stage in momentum life cycle. From table IX of Lee and Swaminathan (2000), the return on equity does improve over the next three years for low volume winner, which shows that the investors in this stage do have accurate perception about the future. From Figure 3 of Hvidkjaer (2001), the buying pressures from both large trades and small trades in this period trend up gradually, signaling the gradual diffusion of information. The buying pressure from large trades are higher than the buying pressure from small trades, which shows that large traders as a group are better informed than small traders.

As the technology goes through various stages from R&D to production, the potential becomes clearer to more people. This means that the level of equivocation gradually reduces to more people, which sustains buying interest and share prices increase gradually. As the technology becomes adopted in production and profit figures become public, the level of equivocation decrease further and the pool of investors increases further. Eventually, both the sustained increase of stock price and stable pattern of profit increase, which are very easy to understand by the general public, attract large amount of buyers, which results in high trading volume and pushes the stock prices further up. This corresponds to the high volume winner stage in momentum life cycle. From Figure 2,3 of Hvidkjaer (2001), there is a steady and higher buying pressure among large traders than in low volume winner stage, signaling a consensus of bullish sentiment from informed investors. Because of this consensus, the return of this stage is extremely high. (Lee and Swaminathan, 2000, Table IV) From Table IX of Lee and Swaminathan (2000), the return on equity is very high for high volume winners. However, the high return of the company will attract the attention of not only investors but also competitors, which will try to produce same or similar products for this high profit market.

From Figure 1, the value of some information that is known to everyone is zero. As the good news reaches most investors, the security is probably already fully or over priced. Among the increased pool of investors, more and more investors understand very little of the fundamentals behind the technology and depend on easy to understand signals such as coverage from popular media and stock price movement to make trading decisions. For this group of investors, they will stop buying only when the opinion of public media changes and the trend of price increase reverses significantly. As stock price keeps increasing, momentum trading becomes highly profitable, which will eventually push the share prices higher than the fundamental value. Since large investors spend more resources in investment, they are generally better informed than small investors. As share prices become highly overvalued relative to fundamentals, large investors start to unload positions while small investors keep buying. As the selling pressure from large investors becomes greater than the buying pressure from small investors, the trend of price increase reverses to price decrease. (Hvidkjaer, 2001, Figure 2) This is the period of high volume losers in momentum life cycle. In the period of high volume loser, as competition intensifies, return on equity drops sharply from previous years. (Lee and Swaminathan, 2000, Table IX).

As the pattern of price drop becomes clear, more and more people joined the selling. After large investors and some of the small investors have finished unloading the positions, the volume of trading will decline, which is the period of low volume loser in momentum life cycle. This period is characterized by active selling of small investor. (Hvidkjaer, 2001) Since small investors are typically slow to understand information, their active selling, after the selling by large investors, signals the selling is overextended, which indicates the low volume losers will rebound and earn high future return in general. From the operation point of view, this is the worst time for the company. Overcapacity of a once high profit margin industry pushes down the return on equity further from the high volume loser stage. There are probably some layoffs of labor and write off of capital. But the return on equity will gradually regress toward normal level. (Lee and Swaminathan, 2000, Table IX).

The above paragraphs describe the patterns of information processing and trading when the initial news is positive. When the news is negative, a similar pattern exists at opposite directions. Since there is a cost shorting stocks and there are many institutional constraints on shorting stocks, short selling is much more difficult than buying stocks. With good news, there are many potential buyers. With a bad news, the sellers are largely confined to existing share holders. So overreaction is less strong on bad news. The statistical results, which are the average of all phenomena, mainly reflect the action from good news instead of bad news. With this observation in mind, we can discuss the following:

"The Hong and Stein (1999) model predicts that momentum profit should be larger for stocks with slower information diffusion. If we make the assumption that scarcity of trading leads to insufficient diffusion of information, then the Hong and Stein model would predict a greater momentum effect among low volume stocks. Our result indicate this to be true among winners but not among losers. That is, low volume winners have greater momentum, but low volume losers actually have less momentum." (Lee and Swaminathan, 2000, p. 2062)

The information theory indicates that low trading volume may reflect either a lack of understanding of some new information or a lack of information. From the above discussion, there are two types of low volume losers. The first, which is an average representative from Lee and Swaminathan's statistical results, is part of a cycle that is triggered by some good news. It has experienced the cycle of low volume winner, high volume winner, high volume loser and low volume loser. The low volume loser period represents the end of an information processing cycle. This is why it exhibits less momentum. The other type is the low volume loser period at the beginning of an information processing cycle that is triggered by a bad news. (Hong, Lim and Stein, 2001) These low volume losers do exhibit strong momentum in low rate of return. So the apparent inconsistency of both views can actually be reconciled with a more detailed analysis from information theory.

The above analysis shows that securities often experience cycles of underreaction and overreaction as the result of investors' information processing. What determine the level of underreaction and overreaction? It depends on how much we understand the fundamentals. If the fundamentals are easy to understand by many people, both initial underreaction and eventual overreaction will be small. If the fundamentals are difficult to understand, mispricing can be substantial. We can have a look at glamour stocks. Glamour stocks are from companies with high earning growth. This means they have very few potent competitors, which generally indicates the lack of deep understanding about the particular products or production systems. That is to say, there is a high level of information asymmetry between the companies of glamour stocks and the general public. Initially, these types of companies are underpriced because few people understand them. However, the solid earning growth of these companies makes the share prices grow continuously, generating clear technical signals. The clarity of technical signals and vagueness of fundamental information will eventually cause high level of overreaction. Statistical results show that stocks undergoing price momentum over longer period will exhibit higher level of reversal. (Lee and Swaminathan, 2000, Table I) Economy wide, great speculative bubbles are generally associated with "new era" or "new economy", when the general public is touched by the profound influence of technology breakthroughs while having little understanding of the underlying mechanisms. (Shiller, 2000)

In the following, we will answer the three questions posed by Lee and Swaminathan at the end of their paper.

"First, the asymmetry in the timing of momentum reversals between winners and losers remains a puzzle. We show that low volume losers rebound quickly and outperform high volume losers with the next three to 12 months. However, it takes low volume winners longer (more than 12 months) to significantly outperform high volume winner. We know of no explanation for this timing difference." (Lee and Swaminathan, 2000, p. 2067)

From our analysis of information processing cycle, the low volume winner stage is the gradually understanding of fundamental news about a firm. Since the understanding of fundamentals is very costly and generally take very long time, it will take long time for low volume winners to significantly outperform high volume winners. The high volume loser stage is when large investors are already well informed about the overpricing and are active sellers. The price at this stage is supported by active buying of small investors, which mainly respond to popular media coverage and technical signals. (Hvidkjaer, 2004) Since coverage from popular media and technical signals, which are information with low cost and low value, are easier to understand than details about fundamentals, the price adjustment at this stage is much faster.

"Second, with the possible exception of the disposition effect from the behavioral literature, we know of no explanation for why trading volume should decline when firms fall out of favor."

The volume of trading reflects how many investors believe they can make profitable trades. When stocks are out of favor, few people believe they can make a profit buying these stocks. Hvidkjaer's detailed analysis shows that losing stocks do experience consistent selling pressures over a long period of time. The low volume of trading when firms fall out of favor reflects one fundamental asymmetry in security trading. For a stock, there are always more potential buyers than potential sellers, who are largely existing share holders.

"Finally, we find it remarkable that measures as readily available as past returns and trading volume can have such strong predictive power for returns. … Why this information is not fully reflected in current prices is another puzzle we leave for future research."

From (3), how much information we can understand depends on our background knowledge about the information and how much weight we assign to the information. From the efficient market theory, trading volume carries very little information. So little weight was given to the idea that trading volume might contain valuable information, which inhibited the research on this direction in the past.

*D. On equity premium*

Mehra and Prescott (1985) observed that the large size of risk premium on US equities can not be explained by the standard general equilibrium models and called it a puzzle. Among much research generated by this observation, two approaches are relevant to our study. One approach attributes the high risk premium to loss aversion by investors.

Barberis and Thaler (2003) provided a survey of works along this line. The other is survivorship bias proposed by Jorion and Goetzmann (1999).

From the generalized entropy theory of information, how long a pattern persists depends on the cost of learning. It is often very costly to gain a deep understanding of a company or an industry, especially when the industry is new. The equity premium puzzle, however, is a very simple pattern on financial data, which, once discovered, can be understood very easily by the investment public. The strategy of profiting from the high equity premium is easy to implement and of low risk. This indicates that the pattern of high equity premium, if it does exist, is a short term one.

Whether high equity premium is a pattern or a result of selection bias can be answered by more comprehensive data. Jorion and Goetzmann (1999) documented that among all the equity markets around the world in the past century, the US market had the highest return. They argued that US market had the highest return because US was the most successful economic system in the world in the last century. "For 1921 to 1996, U.S. equities had the highest real return of all countries, at 4.3 percent, versus a median of 0.8 percent for other countries. The high equity premium obtained for U.S. equities appears to be the exception rather than the rule." (Jorion and Goetzmann,1999, p. 953) Their conclusion is consistent with the generalized entropy theory of information.

**IV. The relation with other models of behavioral finance**

Recently, several behavioral models provide frameworks to interpret the short to intermediate term momentum and long term reversal of return. In this section, we will discuss the relation between the generalized entropy theory of information and these models. Daniel, Hirshleifer and Subrahmanyam (1998) explain momentum in terms of both initial and delayed overreaction, while Barberis, Shleifer and Vishny (1998) and Hong and stein (1999) explain momentum in terms of initial underreaction and followed by delayed overreaction.

From the information theory, the absorbing of a new information is a gradual process, in which the equivocation gradually reduces. So stock prices generally underreact to new information initially, which is confirmed by empirical evidences (Hvidkjaer, 2001). This is consistent with the models of Barberis et al. (1998) and Hong and stein (1999). In the following, we will make further analysis of these two models.

Barberis et al. (1998) utilize the concept of conservatism to understand underreaction. Conservatism states that individuals update their beliefs slowly in the face of new information. This property is a natural result from formula (3). Barberis et al. (1998) attribute overreaction to representativeness heuristic. "People rely on a limited number of heuristic principles which reduce the complex tasks of assessing probabilities and predicting values to simpler judgmental operations" (Tversky and Kahneman, 1974, p.1124). Many investors don't want to spend tremendous resource to research fundamentals. They rely on a limited number of heuristic principles, such as technical signals and opinions from popular media, which reduce the complex tasks of assessing probabilities and predicting values of stocks to simpler judgmental operations with low cost. As we analyzed in the last section, this reliance on simple heuristic principles leads to overreaction in the asset market.

Hong and Stein's (1999) results are built on three key assumptions. The first two assumptions are that traders are classified as "newswatchers" and "momentum traders" according to their information processing abilities. They commented that, "the constraints that we put on traders' information-processing abilities are arguably not as well-motivated by the experimental psychology literature as the biases in Barberis et al. (1998) or Daniel et al. (1998), and so may appear to be more ad hoc" (Hong and Stein, 1999, p. 2145). These assumptions can actually be derived naturally from the entropy theory of information. Depending on the value of assets under management, different investors will choose different methods of information gathering with different costs. "Newswatchers" are large investors who are willing to pay a high cost to collect private information and to make a deep understanding of public information. "Momentum traders" are investors who spend less cost or effort on information gathering and rely mainly on easy to understand low cost information such as coverage from popular media and price momentum signals. Cohen, Gompers and Vuolteenaho (2002) show that institutional investors buy on fundamental news while individual investors buy on price trends. The third assumption of Hong and Stein (1999) is that private information diffuses gradually across the newswatcher population. The gradual diffusion of private information means that the number of people who enjoy low level of equivocation on some information gradually increases.

Both the reduction of equivocation of a representative investor and the increase of number of investors who have low level of equivocation on information contribute to the gradual reduction of underreaction, which generates momentum. Both representativeness heuristic and "momentum trader" can generate overreaction, which will lead to eventual reversal. The information theory can further distinguish the models of Barberis et al. (1998) and Hong and stein (1999). In Barberis et al. (1998), a representative investor make trading decisions. In Hong and stein (1999, 2003), investors are heterogeneous. Investor heterogeneity can be understood naturally because of the different background of the investors and different cost that different investors are willing to pay to gather information. Empirical evidences show that investor heterogeneity exists in financial markets and plays an important role in the formation of trading patterns (Hvidkjaer, 2001).

From (3), the understanding of information depends on the background knowledge. Investors take longer time to understand information from sources they are less familiar with. Hong, Lim and Stein (2001) empirically confirm that information from small firms, from firms with low analyst coverage and from firms with bad news, which managers are reluctant to release, generally diffuse slower. From Hvidkjaer (2001), the selling pressures on loser generally are stronger and last much longer than buying pressures on winners, suggesting information processing is less efficient on bad news.

After discussing the existing behavioral models, Lee and Swaminathan summarized, "existing theories of investor behavior do not fully account for all of the evidence. … none of these models incorporate trading volume explicitly and, therefore, they cannot fully explain why trading volume is able to predict the magnitude and persistence of future price momentum" (Lee and Swaminathan, 2000, p. 2066). Trading volume, on the other hand, is an integral part of the model of investment behavior based on the generalized entropy theory of information. This model answers many questions on the gaps between existing theories and empirical evidences.

## V. Conclusion

Barberis and Thaler(2002) commented that, "models typically capture something about investors' beliefs, or their preferences, or the limits of arbitrage, but not all three." In this work, we show that all three can be understood from the law of statistical physics and evolutionary adaptation. People's belief systems are evolved to process information efficiently. People's preferences are evolved to increase the chance of survival in the environment of our evolutionary past. Arbitrage is limited by the cost of information, which is in general an increasing function of the value of the information. We show that this information theory based model of investor behavior captures the detailed patterns in asset market.

**Figure Captions**

Figure 1: Information value and probability

Figure 2: Overnight rate of return and trading volume of WestJet stock
surrounding the date when Jetsgo announced bankruptcy

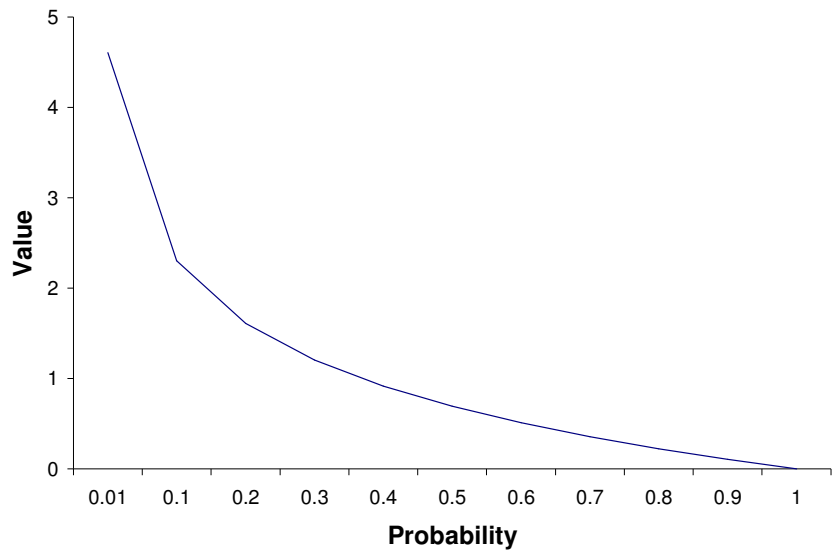

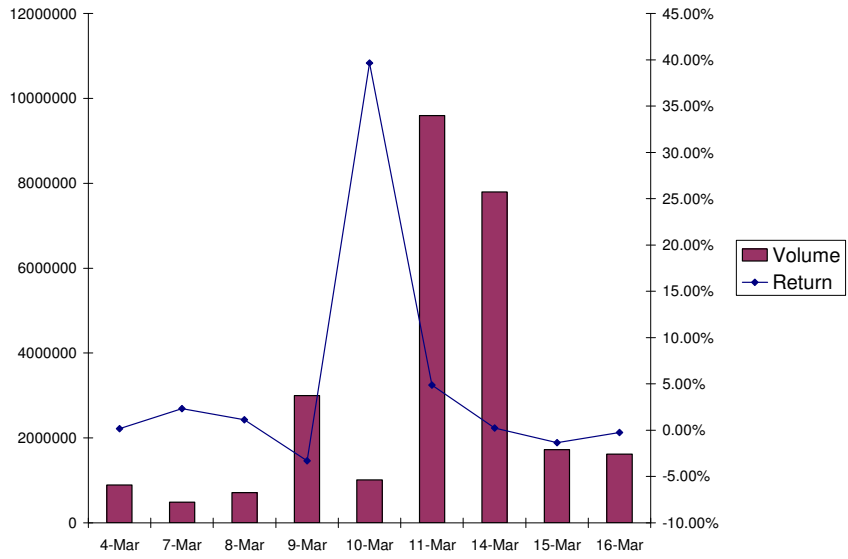